# *Volatile Organic Compounds for Stress Detection: A Scoping Review and Exploratory Feasibility Study with Low-Cost Sensors*

Nicolai Plintz
Learning, Design & Technology
University of Mannheim
Mannheim, Germany

Marcus Vetter
Institute for Applied Artificial Intelligence
and Robotics (A²IR)
Technische Hochschule Mannheim, Germany

Dirk Ifenthaler
Learning, Desing & Technology
University of Mannheim & Curtis University
Mannheim, Germany & Perth, Australia

*Abstract* - Volatile organic compounds (VOCs) represent a novel but underexplored modality for emotion recognition. This paper presents a systematic evidence synthesis and exploratory investigation of VOC-based affective computing using low-cost sensors. Study 1, a systematic scoping review following PRISMA-ScR guidelines, analyzed 16 studies from 610 records across breath, sweat, skin, and urine biosources. Evidence indicates that stress and affective states are reflected in VOC signatures (aldehydes, ketones, fatty acids, sulfur compounds), though with considerable heterogeneity. Current research relies predominantly on laboratory-grade GC-MS or PTR-MS, while wearable sensors provide pattern-level outputs without compound-specific identification - a critical gap for practical systems. Study 2 (n=25) investigated whether low-cost TVOC sensors (BME688, ENS160) combined with physiological monitoring (HR, HRV, GSR) can detect laboratory-induced stress. Exploratory analysis revealed that high cardiovascular reactors exhibited elevated TVOC during arithmetic stress (d=1.38), though requiring replication in larger samples. Substantial interindividual variability emerged (CV>80%), with coupling patterns moderated by baseline emission levels and temporal lags of 30-80 seconds. Random Forest-based multimodal classification achieved 77.3% accuracy (5-fold CV). SHAP analysis indicated VOC sensors contributed 24.9% of model performance. Leave-one-subject-out validation yielded 65.3% accuracy, highlighting the need for individual calibration. This work provides three contributions: (1) comprehensive mapping of VOC biomarker evidence and technological gaps, (2) initial demonstration that low-cost sensors can capture stress-related VOC patterns in multimodal fusion, and (3) identification of key implementation challenges. Findings require replication in larger samples (n>=50).

I. INTRODUCTION

The ability to detect human emotions and stress states has long fascinated researchers, with evidence suggesting that even animals possess remarkable capabilities in this domain. Dogs, with their exceptional olfactory sensitivity, have demonstrated the ability to detect human emotional states through scent-based cues. The study by Wilson and colleagues [1] provides strong initial evidence that dogs can distinguish between breath samples from stressed and relaxed individuals with high accuracy (93.75%), suggesting that psychological stress causes measurable changes in volatile organic compounds (VOCs) in breath [1]. This canine ability highlights the potential of chemical signatures as indicators of human emotional states and opens avenues for technological applications. Building on these insights from nature, emotions might now be detected and assessed through various computer-based measurement methods. Current multimodal emotion recognition systems primarily rely on facial expression analysis, speech emotion recognition, and physiological signal processing [2]. Traditional approaches to emotion recognition, such as facial expression analysis, speech processing, or physiological sensing, often face practical limitations, for instance, due to poor lighting conditions, the absence of verbal communication, or the need for direct skin contact (e.g. [3]). This has motivated the search for complementary modalities. One promising direction is the analysis of volatile organic compounds (VOCs), which may reflect emotional states through underlying metabolic processes (e.g., [4], [5]). VOCs are low-molecular-weight substances that rapidly evaporate or sublimate, making them theoretically detectable in the surrounding air [6]. They can be measured for emotion-related research using compounds in sweat (e.g., [5], [7]), skin gases (e.g., [8] - [11]), breath (e.g., [12], [13]), or, less frequently, from urinary VOCs [14]. Early studies suggest that VOCs could offer certain advantages: they can be measured without requiring physical contact, for example, through electronic nose (e-nose) devices that detect VOC patterns in exhaled breath or skin emissions [13], [15]. Moreover, VOC-based sensing operates largely outside of conscious awareness and is difficult to control deliberately. The physiological basis for VOC-based emotion detection is tentatively supported by preliminary research on emotionally induced hyperhidrosis [16]. Preliminary evidence suggests that certain affective states, particularly fear, may be detectable via sweat-derived VOCs, with a possible dose-response relation: stronger fear in the donor is associated with more axillary sweat, and higher sweat amounts with greater emission of volatile molecules [17]. These VOCs can be detected in humans through various analytical methods such as gas chromatography-based approaches (e.g. [5], [8], [13]), electronic nose systems (e.g., [12], [18], [19]), or mass spectrometry methods [20], [6]. It is important to note that while GC-MS studies have identified individual VOC biomarkers, wearable devices and e-nose systems typically capture broader VOC patterns that require machine learning for classification. However, VOC research has traditionally been conducted within medical and analytical chemistry domains, focusing on disease diagnosis rather than emotional state detection (e.g. [21]). This creates critical research gaps for affective computing: the field lacks standardized protocols for emotion-focused VOC analysis, existing research has not been integrated into affective computing frameworks, and the technical feasibility for real-world deployment remains unclear. Linking VOC biomarkers to specific affective states such as stress (e.g., [18], [9]) or discrete emotions [22], [5] remains challenging and requires systematic investigation. Moreover, the current evidence base is inconsistent, often limited to small samples and laboratory settings, and validated biomarkers for real-world applications remain to be established [23]. For affective computing applications, the transition from laboratory-based VOC analysis to practical wearable systems requires a systematic assessment of three key questions:

(1) Which VOC biomarkers have been identified for emotional states and what is the consistency of findings across studies?

(2) What is the technological feasibility of these biomarkers for integration into wearable systems?

(3) Can low-cost sensors detect stress-related VOC changes in combination with established physiological measures?

To address these questions, we conducted a systematic scoping review and exploratory empirical investigation. Study 1 comprises a systematic review following largely PRISMA-ScR guidelines [24] to map existing VOC biomarker evidence (RQ1) and assess technological feasibility (RQ2). Study 2 examines whether low-cost TVOC sensors (BME688, ENS160) combined with physiological monitoring (HR, HRV, GSR) can detect laboratory-induced stress. Specifically, we investigate: (RQ3) whether stress-induced TVOC changes correlate with physiological responses and whether cardiovascular reactivity moderates this relationship; (RQ4) the degree of interindividual variability in physiological-VOC coupling patterns; (RQ5) classification accuracy through multimodal sensor fusion and feature importance via SHAP analysis; and (RQ6) generalization to unseen individuals through leave-one-subject-out validation. This work makes three contributions to affective computing research. First, we provide a comprehensive synthesis of VOC biomarker evidence across biosources and measurement approaches, revealing critical gaps between laboratory identification and wearable implementation. Second, we present initial exploratory evidence that low-cost TVOC sensors may capture stress-related patterns in laboratory settings. Third, we identify key implementation challenges, including substantial interindividual variability (CV > 80%) and the need for individual calibration, which provide a foundation for future research. Given the nascent state of VOC-based affective computing, Study 2 employs an exploratory design (n=25) focused on hypothesis generation rather than definitive conclusions, with findings requiring replication in larger samples (n≥50).

## II. Study 1: Methodology - Systematic Scoping Review

To answer the first two research questions, this study employed a systematic scoping review, conducted largely in accordance with the PRISMA-ScR guidelines [24], to map the existing literature on VOC-based emotion recognition. Four databases were searched (IEEE, ACM, Web of Science, PubMed) using the search string: ("volatile organic compounds" OR VOC OR "electronic nose" OR "e-nose") AND (stress OR emotion* OR affect OR happiness OR sadness OR anger OR fear OR surprise OR disgust).

Included were studies investigating VOCs as biomarkers of human emotional states, employing technological detection approaches, published in English (2020–2025), presenting original empirical research. The initial search yielded 610 records. After removing 20 duplicates, 590 records were screened based on title and abstract. Of these, 45 articles underwent full-text assessment, and 16 met all inclusion criteria (see PRISMA flow diagram, Fig. 1). Data extraction focused on measurement methods, target emotions, sample sizes, algorithms, performance metrics, and identified VOC biomarkers.

Fig. 1. Flow-Chart

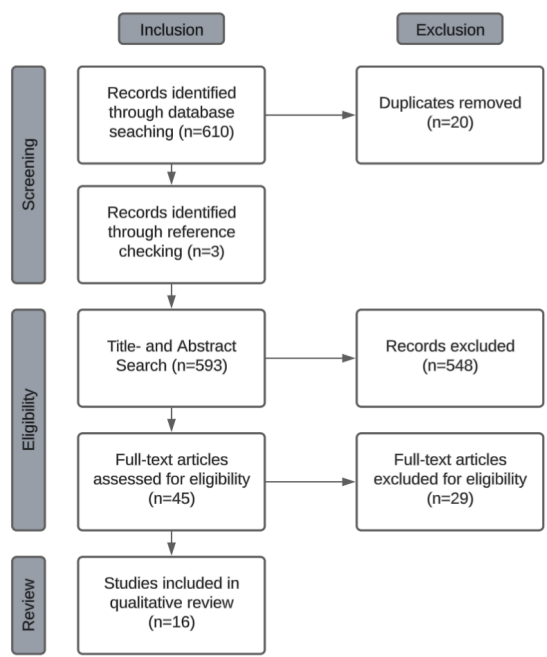

## III. STUDY 1: Results: Systematic Scoping Review

### A. What volatile organic compounds have been identified as biomarkers for human emotional states, and what is the consistency of findings across different measurement approaches?

Only a small number of studies (n = 12) have directly investigated affective states or discrete emotions via VOCs, often with limited participant samples (e.g., [22], [8]). The available findings can broadly be grouped into breath- and sweat-related detection approaches, as summarized in Table 1. Studies focusing on exhaled breath have so far yielded heterogeneous and mostly preliminary results. Yu and Xu [22] tested a low-cost setup in a small pilot study (n = 5) but did not identify any specific VOC biomarkers associated with fear, happiness, or disgust. In contrast, a large-scale study by Xu et al. [12] (n = 1,207) used an electronic nose on breath samples and also found no generalizable signatures, as machine learning models overfitted during training but showed no predictive power on test data. Wicker and colleagues [6] reported statistical associations between certain mass peaks (e.g., m/z 18.0338) and film scenes labeled "blood (violence)," but emphasized the exploratory nature of these findings. In a controlled study of sexual arousal (n = 24), Wang et al. [20] observed decreases in $CO_2$ and isoprene, as well as occasional increases in compounds such as phenol, cresol, and indoles. However, these effects were observed only in a few individuals. Finally, Zhang and colleagues [13] found that cognitive fatigue was associated with a significant increase in isoprene, while changes in acetone, ethanol, and limonene did not reach statistical significance. Preliminary findings from sweat and skin emissions suggest potentially distinct biomarker profiles across sampling regions, though the evidence remains inconsistent. In axillary sweat, Smeets and colleagues [5] distinguished fear from neutral states ($Q^2 = 0.85$) based on aldehydes and ketones, while neutral sweat contained more esters and cyclic molecules; happiness showed a heterogeneous pattern. Tungkijanansin et al. [15] tentatively identified seven stress-related compounds (ammonia, diethyl ether, methanol, octane, pentane, acetone, dimethylamine) in a large sample (n = 154) using GC-IMS, but marker specificity remains uncertain. Wang et al. [25] analyzed body odour from worn T-shirts and identified multiple fatty acids and alcohols, though no single compound was consistently linked to emotion. In forehead skin emissions, Martin et al. [8] reported benzoic acid and n-decanoic acid as elevated under stress, and decreases in a xylene isomer and 3-carene. Similar candidates were summarized by Zamkah [10] in a review, while also highlighting strong methodological heterogeneity across studies. Lucchi et al. [9] identified 17 stress-associated VOCs, including 3-methylpentadecane, 2-hydroxyethylacetate, and several alkanes. Acevedo [18], [19] achieved high stress classification with forehead-mounted e-nose sensors, but without compound-level identification. For general skin gases, Katsuyama and team [11] found that stressed participants reproducibly emitted allyl mercaptan and dimethyl trisulfide, which correlated with sympathetic activity (r = 0.66, p < 0.01) and elicited anxiety, confusion, and fatigue in perceivers. Gioao and colleagues [7] reported that dodecanoic acid was a potential stress marker during Stroop tasks, significantly predicting changes in HRV and nasal temperature.

Besides breath- and sweat-derived VOCs, only one study has focused on urinary VOCs, indicating that evidence from this biosource is very limited, particularly with respect to affective states such as stress. He and colleagues [14] analyzed urinary metabolites of VOCs (mVOCs) in a large population sample (n = 6,966) and identified N-Acetyl-S-(2-cyanoethyl)-L-cysteine (CYMA) as the only metabolite consistently associated with suicidal ideation across all models. Other investigated metabolites (e.g., HPMMA, 3-HPMA, MHBMA3, AAMA) showed no stable associations. While not directly linked to discrete emotions, these findings suggest that certain urinary mVOCs may provide indirect indicators of negative affective states.

TABLE I
Results from the systematic scoping review

| No. | Study | Biosource | Key findings | Method / Sensor | Target affective states | N | Biomarkers | Advantages | Limitations |
|---|---|---|---|---|---|---|---|---|---|
| 1 | [22] | Exhaled breath + saliva | No specific VOC biomarkers identified; BAC and saliva pH showed no correlation with emotions | Saliva pH test, BAC test, Apple Watch HR | Happiness, fear, disgust | n = 5 | – | Low-cost, feasible | No correlation; very small sample |
| 2 | [12] | Exhaled breath | No generalizable VOC signatures; ML models overfitted and failed in test data | Electronic nose (Cyranose 320 with 32 sensors) | Depression, anxiety, substance use disorders | n = 1.207 | – | Large sample; ML approach | No predictive power; $R^2 \approx 0$ |
| 3 | [6] | Exhaled breath + skin emissions | Statistical associations between mass peaks (e.g., m/z 18.0338) and film scenes, but exploratory only | PTR-ToF-MS, $CO_2$ detector | Fear (explorative, film-induced) | – | m/z 18.0338 (explorative, might belong to protenated ammonia) | Rich dataset, annotated film scenes | No validated emotional biomarkers |
| 4 | [20] | Exhaled breath | Decrease of $CO_2$ and isoprene during sexual arousal; occasional increases in phenol, cresol, indoles (few responders) | PTR-ToF-MS, CRDS | Sexual arousal, positive/negative arousal | n = 24 | Isoprene and $CO_2$ (decrease for both genders), phenol/cresol/indoles (increase for some male), dopamine-related and cresol (increased for female) | High temporal resolution | Effects limited to outliers |
| 5 | [13] | Exhaled breath | Cognitive fatigue associated with significant isoprene increase; other VOCs not significant | GC-MS (Agilent 6890–5973N) combined with secondary thermal desorption system (ATD-650) | Cognitive fatigue (N-back task) | n = 30 | A total of 15 VOC were detected, just an increase in Isoprene was significant; acetone, ethanol, limonene had changes but not significant | Clear task design | Only one significant biomarker significant |
| 6 | [5] | Axillary sweat | Fear vs. neutral discriminable ($Q^2 = 0.85$, significant); Happy vs. neutral moderate ($Q^2 = 0.6$; significant); Fear vs. Happy weak ($Q^2 = 0.33$, significant); happiness heterogeneous | HSSE + GC×GC-TOFMS | Fear, happiness, neutral | n = 24 | Aldehydes/ketones (fear); esters/cyclics (neutral) | Clear biomarker profiles | Small, homogeneous sample |
| 7 | [15] | Axillary sweat | Seven tentative VOCs distinguished high vs. low stress; classification accuracy up to 89% | HS-GC-IMS + E-nose (8 sensors) | Stress (high vs. low) | n = 154 | Ammonia, diethyl ether, methanol, octane, pentane, acetone, dimethylamine | Non-invasive, rapid screening (5-30 min), dual-method validation, high accuracy (87-89%) | Tentative IDs; deodorant confounds; only n = 46 screened (14 high stress, 32 controls; screened from 154 nurses) |
| 8 | [25] | Sweat/skin (T-shirt samples) | Body odor increased N-back accuracy but decreased emotional valence; $CO_2$ (800-3500 ppm) and body odor had no significant effects on driving performance; no stress-specific VOC markers identified | GC-MS of worn T-shirts (axilla, chest, back) | Not primary on emotion: Emotion (SAM: valence, arousal, dominance), sleepiness (SSS), task load (NASA-TLX) | n = 25 | 26 VOCs identified (multiple fatty acids, alcohols, aldehydes); 12 common in both sexes; no stress-specific markers validated | Multi-method approach (GC-MS + behavioural tests + questionnaires) | No specific stress markers; small sample size; no quantitative VOC measurement in cabin air |
| 9 | [8] | Forehead sweat | Stress ↑ benzoic acid, n-decanoic acid; stress ↓ xylene isomer, 3-carene | TD-GC-MS, PDMS coupons | Stress (vs. neutral) | n = 15 | Benzoic acid ↑, n-decanoic acid ↑, xylene isomer ↓, 3-carene ↓ | Clear candidate VOCs | Small sample |
| 10 | [9] | Forehead skin emissions | Identified 17 stress-associated VOCs (e.g., alkanes, 3-methylpentadecane, 2-hydroxyethylacetate) | GC-MS, Sorb-Star® polymer | Stress (cognitive tasks) | n = 35 (female) | 17 candidate VOCs | Robust within-subject design | Needs replication; only women |
| 11 | [18] | Forehead skin | Stress vs. relaxation classification accuracy ~90% with E-nose sensors; no specific VOCs identified | MOX gas sensors (8 types) + E-nose | Stress vs. relaxation | n = 25 | – | High accuracy with ML | Unspecific sensors; no biomarkers |
| 12 | [19] | Forehead sweat | Improved classification accuracy (96%); same limitations as 2021a | MOX sensors (same as 2021a) | Stress vs. relaxation | n = 25 | – | Slightly improved performance | No biomarker identification |
| 13 | [11] | Skin gases | Stress → emission of allyl mercaptan & dimethyl trisulfide; correlated with sympathetic activity; perceived as "stress smell" inducing anxiety, confusion, fatigue in others | Skin gas sampling + ECG + saliva | Psychological stress (Trierer Stress Test) | n = 40 + replication subsamples | Allyl mercaptan, dimethyl trisulfide | Clear, reproducible stress markers | Limited generalizability |
| 14 | [7] | Sweat (axilla & back) | Dodecanoic acid predicted stress response; correlated with HRV and nasal temperature | HS-GC-MS, physiological signals | Stress (Stroop task) | n = 10 | Dodecanoic acid | Multi-modal data | Very small sample |
| 15 | [10] | Sweat + skin gases | Review identified candidate biomarkers from literature; profiles varied by body site and population | Systematic review | Stress | – | Benzoic acid, n-decanoic acid, xylene isomer, 3-carene, acetophenone, benzothiazole etc. | Broad synthesis | Heterogeneous, inconsistent findings |
| 16 | [14] | Urine (mVOCs) | CYMA consistently associated with suicidal ideation; other metabolites not stable | Elastic net regression of urinary mVOCs | Negative affect (suicidal ideation) | n = 6,966 | CYMA (N-Acetyl-S-(2-cyanoethyl)-L-cysteine) | Very large sample | Not linked to discrete emotions |

*B. What is the technological feasibility of identified VOC biomarkers for integration into wearable emotion recognition systems, and what sensor technologies are required for implementation?*

Regarding technological feasibility, our review indicates that wearable systems for VOC-based emotion recognition remain largely unexplored. Most studies continue to rely on laboratory-based techniques such as GC-MS or PTR-MS (e.g., [13], [25]), whereas portable devices remain limited. Current wearable chemosensing platforms, particularly those based on MOX sensors, provide pattern-level readouts rather than compound-specific quantification [26], [18]. While recent work indicates that the sensitivity and accuracy of these sensors are improving [27], performance gains are achieved mainly at the pattern level rather than through validated single biomarkers. There is also a trend towards more compact sensor architectures, where deep learning enables quasi-analytical discrimination and even concentration estimation [28]. Nevertheless, these advances do not yet match the compound-level resolution of laboratory methods, highlighting a gap between biomarker identification under controlled conditions and their implementation in wearable affective computing systems.

*C. Study 1: Discussion and Implications for Affective Computing – Systematic Scoping Review*

Our systematic scoping review highlights that VOC-based biomarkers provide some initial indications that stress and, in some instances, emotional states might be reflected in chemical signatures. While these early findings are encouraging, the overall evidence remains inconsistent and limited to small or highly controlled studies. At the same time, sensor technologies – particularly wearable systems – are not yet sufficiently advanced to capture such signals in real-world settings reliably. Current devices typically provide only general pattern-level information rather than validated compound-specific readouts, underscoring the gap between laboratory findings and practical applications. To move the field forward, future research should combine advances in wearable sensor technology with broader data collection efforts. Larger datasets of VOC signals from wearables will be essential to establish more robust evidence, particularly when integrated with other modalities through approaches such as cross-modal learning [29]. Linking wearable VOC measurements with established gold standards – for instance, EEG, which can already achieves accuracies above 90% in emotion recognition [30] – may help clarify the added value of VOCs and provide a stronger basis for understanding their role in affective computing.

In particular, our review points to three key implications for the Affective Computing community. First, multimodal integration is essential: VOC data should not be evaluated in isolation but in combination with established modalities such as EEG, EDA, HRV, facial expressions, or speech. Only through such integration can the added value of VOCs be assessed and risks of overfitting avoided. Second, there is a strong need for benchmarking and datasets. To date, no open benchmark datasets exist that provide synchronized VOC and emotion measurements. The availability of such multimodal corpora, including VOCs, would significantly accelerate progress in this domain. Third, standardization of data collection protocols is critical. Clear guidelines regarding sampling sites, duration, task designs, and controls for confounding factors (e.g., food intake, deodorant use, environmental conditions) are required to ensure comparability and reproducibility across studies.

IV. STUDY 2: METHODOLOGY - EXPLORATORY EMPIRICAL STUDY

Study 1 revealed that while GC-MS and PTR-MS provide compound-level biomarker identification, their laboratory-scale instrumentation precludes integration into wearable systems. To bridge this gap, we conducted an exploratory empirical investigation using low-cost TVOC sensors (BME688, ENS160) with physiological monitoring (HR, HRV, GSR) during laboratory stress induction. This investigation comprises two complementary analyses of a single Dataset:

Study 2a: Physiological-VOC Coupling Analysis (addresses RQ3-RQ4) Examines whether stress-induced TVOC changes correlate with cardiovascular/electrodermal responses and quantifies inter-individual variability.

Study 2b: Multimodal Classification Performance (addresses RQ5-RQ6) evaluates machine learning algorithms for binary stress detection and assesses generalization to unseen individuals.

The following sections describe the shared methodology, followed by separate results and discussion sections for each analysis.

*A. Participants*

Twenty-five participants ($N = 25$; 18 male, 7 female) aged 23-61 years ($M = 29.04$, $SD = 7.52$) from eight countries were recruited for this exploratory feasibility study. All participants provided informed consent prior to participation and received no financial compensation. This sample size reflects the exploratory, proof-of-concept nature of the investigation. We acknowledge the limited sample as inherent to feasibility studies with custom sensor setups.

*B. Experimental Setup*

Measurements were conducted in three different laboratory rooms, including a university research laboratory, a university of applied sciences laboratory, and a corporate office space. All rooms underwent standardized ventilation protocols, with thorough ventilation prior to each session to minimize ambient VOC contamination. Environmental conditions remained stable across measurement sessions. Temperature averaged 22.7°C ($SD = 1.9$, range: 15.9-26.2°C), humidity 40.2% ($SD = 8.6$, range: 23.4-61.8%), and atmospheric pressure 1001.4 hPa ($SD = 7.9$). Ambient TVOC baselines ranged from 8 to 689 ppb ($M = 155$, $SD = 186$). Kruskal-Wallis tests revealed no significant differences across measurement times for temperature ($H = $

9.61, $p = .38$), humidity ($H = 16.19$, $p = .06$), or pressure ($H = 4.71$, $p = .86$), confirming that environmental variability did not systematically confound experimental phases. Participants were instructed to avoid alcohol (24 hours prior), food intake (3 hours), smoking (24 hours), carbonated beverages (3 hours), and strong fragrances, including perfume and deodorant before participation. We employed a custom-built, low-cost sensor setup comprising multiple synchronized measurement devices. Cardiac monitoring was performed using a Polar H10 chest strap for continuous heart rate (HR) and heart rate variability (HRV) measurement, with a sampling rate of 1 Hz for HR and approximately 180 Hz for RR interval recording. Galvanic skin response (GSR) was measured using a custom Arduino-based sensor with finger electrodes placed on the index and middle fingers, sampled at 0.2 Hz.

VOC measurement employed a dual-sensor configuration positioned 5-6 cm from participants' mouths to capture exhaled breath. The first sensor, a BME688, is a multi-temperature gas sensor that measures resistance at 250°C, 320°C, and 400°C, along with environmental parameters such as temperature, humidity, and pressure. The second sensor combined an ENS160 with an AHT21 module (ScioSense) to measure the Air Quality Index (AQI), total volatile organic compounds (TVOC, in parts per billion), equivalent $CO_2$ ($eCO_2$, in parts per million), temperature, and humidity. Both VOC sensors were interfaced via an ESP32 microcontroller at a sampling rate of 0.2 Hz, with 3-second measurement intervals, 150ms heating cycles per temperature step, and 250ms stabilization delays between temperature changes. Primary analyses focused on ENS160 TVOC measurements due to their broader spectral sensitivity and more stable baseline characteristics compared to the temperature-cycled BME688 resistance values, as a pilot attempt to explore neural correlates of stress-induced. VOC changes and electroencephalography (EEG) were recorded using a BrainVision Recorder for two participants. However, equipment failure precluded further data collection, and EEG analysis was excluded from the present investigation. Future work should systematically integrate neural measurements to validate mechanisms of central-peripheral VOC coupling. All sensors were synchronized via event markers logged through a Flask-based web application with millisecond-precision timestamps. The application automatically triggered sensor state changes between baseline and experiment phases via serial commands to the ESP32 microcontroller.

The system logged 19 parameters per measurement cycle, including HR (beats per minute), RR intervals (milliseconds), GSR raw values, gas resistance at three temperatures (kilohms), AQI, TVOC (ppb), $eCO_2$ (ppm), temperature (degrees Celsius), humidity (percent), atmospheric pressure (hectopascals), and within-participant normalized values (Gas320_Norm, TVOC_Norm, GSR_Norm). Due to sensor failures and protocol adjustments during data collection, complete datasets were unavailable for some participants. GSR data were collected from 19 participants, with six excluded due to sensor introduction after initial sessions ($n = 5$) or insufficient skin conductivity ($n = 1$). HR and HRV data were available for 24 participants, with one participant excluded due to a loss of Polar H10 connection during measurement. EEG data were collected from only 2 participants before equipment failure.

*C. Stress Induction Protocol*

We developed a multi-stage stress induction protocol combining elements from the Stroop Color-Word Test [31] and the Trier Social Stress Test (TSST) arithmetic component [32]. Phase 1 (approximately 15 minutes) served as a sensor warm-up and preparation period during which participants completed informed consent procedures and questionnaires while automatic BME688 baseline calibration was initiated. No physiological features were extracted from this phase.

Phase 2 consisted of a 3-minute baseline period during which participants, in a relaxed state, passively viewed and called out animal names displayed on screen.

Phase 3 comprised a modified Stroop Test lasting approximately 200 seconds. Participants completed a computerized Stroop task in either German or English, depending on their native language, in which color words appeared in incongruent ink colors. Participants verbally identified the ink color while the display auto-advanced every second. Additional cognitive load was introduced by requiring participants to recall the last three colors presented. Social-evaluative stress was induced by informing participants that their performance would be compared with others'.

Phase 4 involved a mental arithmetic task lasting 4 minutes. Following TSST protocols, participants performed serial subtraction, starting at 1022 and subtracting 13 repeatedly, under time pressure with multiple stressors. Participants were given a performance target of reaching below 900 within 4 minutes, with the caveat that failure to meet this target would result in their data being unusable. The task automatically restarted upon calculation errors or response delays exceeding 3 seconds. Verbal feedback of "Please concentrate" was provided after errors, and time-pressure announcements were given at the 2-minute and 1-minute marks, emphasizing the need to reach the target value.

Phases 5-7 consisted of a 6-minute recovery period divided into three sub-phases to assess stress recovery. Phase 5 (2 minutes) required participants to count backward aloud from 100 to 50. Phase 6 (2 minutes) consisted of silent rest without speaking. Phase 7 (2 minutes) required participants to count backward aloud from 50 to 0.

*D. Self-Report Measures*

**Demographics and Confounds.** Prior to baseline measurement, participants completed a comprehensive questionnaire assessing potential confounding variables. Demographic information included age, gender, and country of origin. Sleep-related questions assessed the quality rating and duration for the previous night. Recent substance use questions covered alcohol consumption (24 hours), caffeine consumption, and tobacco use (24 hours, 6 months, and regular use patterns). Medication questions addressed psychiatric medications, sedatives, and other prescriptions. Health status questions covered cardiovascular conditions, respiratory infections, and psychiatric diagnoses, including depression and anxiety. Pre-session behavior questions assessed food intake (3 hours), physically demanding activities, and fragrance or cosmetic use, including perfume, deodorant, and body lotion.

**Stress Assessment.** Self-reported stress was measured using the Perkhofer Stress Scale, a pictorial scale combining elements from Visual Analog Scales (VAS) and Faces Rating Scales (FRS) [33]. The scale presents six facial expressions ranging from a broad smile labeled "no stress" to an intensely distressed face labeled "maximum stress". Participants indicated their current momentary stress level at three timepoints: post-baseline (T1), post-Stroop (T2), and post-arithmetic (T3).

### E. Procedure

The experimental session followed a standardized sequence and lasted approximately 30 minutes. The session began with a 15-minute sensor warm-up period during which automatic BME688 baseline calibration was initiated via the Flask application. This was followed by informed consent procedures and questionnaire completion lasting approximately 10 minutes. The baseline measurement consisted of 3 minutes of passive reading of animal names on screen. Following baseline, participants completed stress assessment T1 using the Perkhofer Scale, after which the VOC experiment phase was automatically activated via a Flask-to-ESP32 serial command. Participants then completed the Stroop task, lasting approximately 3.5 minutes, followed by the T2 stress assessment. The mental arithmetic task lasted 4 minutes and was followed by stress assessment T3. The session concluded with three recovery phases totaling 6 minutes (2 minutes each), after which a session completion command was sent to stop VOC data collection. All phase transitions and stress assessments were automatically logged by the Flask application with millisecond-precision timestamps, enabling precise multi-sensor data alignment.

## V. STUDY 2A: PHYSIOLOGICAL-VOC COUPLING ANALYSIS

### A. Data Processing for Coupling Analysis

Raw sensor data were visually inspected for artifacts and sensor failures. Time-series data were aligned using logged event markers from the Flask application. For each participant, baseline-normalized values were computed using the formula:

$$Norm_X = \left(\frac{X_{baseline_{mean}} - X_{experiment}}{X_{baseline_{mean}}}\right)$$

where X represents Gas320 (kilohms), TVOC (parts per billion), or GSR (raw analog-to-digital converter values). This within-subject normalization accounts for individual baseline differences in physiology and VOC emission profiles. Positive normalized values indicate decreases from baseline, such as reduced gas sensor resistance due to increased VOC exposure to the sensor. Cardiovascular reactivity grouping was performed by classifying participants as high or low cardiovascular reactors based on a median split of HR change from baseline to the stress phase. Phase effects were tested using a repeated-measures ANOVA to examine differences across the experimental phases (Baseline, Stroop, Arithmetic, Recovery) within each reactivity group. Individual coupling analysis assessed physiological-VOC coupling through time-lagged cross-correlations with a lag range of 0 to 120 seconds, examining HR-to-TVOC and GSR-to-TVOC relationships for each participant. Moderator analysis employed independent samples t-tests to compare coupling direction (positive versus negative correlations) between high and low TVOC emitters, defined by a split at TVOC_norm = .4 during stress. The significance threshold was set at $\alpha$ = .05, and effect sizes were reported as Cohen's d. EDA was recorded using an analog GSR sensor (Grove) that measures skin resistance (R, in arbitrary units). For analysis and visualization, raw resistance values were reciprocally transformed to skin conductance equivalents (C $\propto$ 1/R) to align with psychophysiological convention, where increased sympathetic activation manifests as increased signal amplitude [34]. This transformation was applied as: C = k/R, where k is a scaling constant. Participants showing significant physiological-VOC coupling ($p < .05$ in time-lagged cross-correlations) were classified as "responders" for subsequent moderator analyses.

### B. Results: Manipulation Check

Manipulation check provided evidence of successful stress induction: Heart rate increased significantly from baseline ($M$ = 73.0 bpm, $SD$ = 9.6) to stress ($M$ = 84.2 bpm, $SD$ = 9.9; $t(23)$ = 9.68, $p < .001$, $d$ = 5.20), with all participants showing individual-level increases ($p < .05$). Electrodermal activity (skin conductance) showed the expected directional increase during stress (M_baseline = 2104 au, M_stress = 4883 au, $\Delta$ = +132%), with 90% of participants (18/19) exhibiting individual-level increases. However, this effect did not reach statistical significance at the group level (p=0.21), likely due to substantial inter-individual variability in baseline skin conductance ($SD$ = 9143 au).

### C. Results: RQ3 - Cardiovascular Reactivity and TVOC Changes

To examine whether cardiovascular stress reactivity modulates VOC emission patterns, participants were classified via median split based on heart rate change from baseline to stress phases. This classification yielded two groups: Low Reactors ($n$ = 12) showing heart rate increases of 2.2 to 9.4 beats per minute ($M$ = 5.74, $SD$ = 2.1), and High Reactors ($n$ = 12) showing increases of 10.6 to 26.4 beats per minute ($M$ = 14.37, $SD$ = 4.8) Exploratory repeated measures ANOVA revealed a significant phase effect for TVOC emissions in High Reactors, $F(6,60) = 4.77, p < .001$, though this finding requires replication given the small subgroup size ($n$ = 12). Post-hoc pairwise comparisons with Bonferroni correction showed that High Reactors exhibited significantly elevated TVOC levels during the arithmetic stress task (Phase 4) compared to baseline (Phase

2), with a mean difference of +0.65 ($p = .022$) and a large effect size ($d = 1.38$). In contrast, Low Reactors showed no significant phase-related changes in TVOC emissions, $F(6,66) = 1.23$, $p = .29$. Despite this group-level effect, within-subject correlations between momentary heart rate and TVOC fluctuations among High Reactors were not significant ($r = .086$, $p = .697$). This dissociation suggests that while individuals with high cardiovascular reactivity show elevated VOC emissions during stress at the trait level, moment-to-moment physiological fluctuations do not reliably predict concurrent VOC changes within individuals.

*D. Results: RQ4 - Interindividual Variability in Coupling Patterns*

Time-lagged cross-correlation analyses revealed substantial interindividual heterogeneity in physiological-VOC coupling patterns. We report uncorrected p-values (α = .05) to maximize sensitivity for hypothesis generation in this exploratory investigation. Of 22 participants with valid heart rate data, 7 (32%) showed significant HR→TVOC correlations at p<0.05. Among 18 participants with valid GSR data, 7 (39%) demonstrated significant GSR→TVOC coupling. Respiratory rate showed the highest responder rate, with 12 of 23 participants (52%) exhibiting significant RR → TVOC correlations. Among participants with significant coupling across all three physiological modalities, 16 of 22 (73%) qualified as responders. However, this responder group was evenly split in coupling direction: eight participants showed positive correlations (higher physiological arousal associated with higher VOC emissions) while eight showed negative correlations (higher physiological arousal associated with lower VOC emissions). The coefficient of variation for correlation strengths exceeded 80% across participants, indicating substantial between-person variability in the magnitude of physiological-VOC coupling even among those classified as responders.

Optimal temporal lags varied across individuals, ranging from 30 to 80 seconds, with no consistent pattern across modalities. This heterogeneity indicates that the time course of physiological-VOC coupling is highly individualized and cannot be characterized by a single universal temporal relationship. To investigate potential moderators of coupling direction, we examined whether baseline TVOC emission levels during stress differentiated participants with positive versus negative physiological-VOC correlations. Among responders showing significant coupling ($p < .05$), participants were classified as High Emitters if their mean normalized TVOC during stress exceeded 0.4 ($n = 6$) or Low Emitters if it fell below this threshold ($n = 6$). High Emitters exhibited mean stress-phase TVOC levels of 0.73 ($SD = .20$), significantly higher than Low Emitters, who showed levels of .06 ($SD = .27$), $t(10) = 4.65$, $p = .002$, $d = 3.0$. Critically, High Emitters demonstrated positive physiological-VOC coupling, where elevated heart rate and slower respiration were associated with increased VOC emissions. In contrast, Low Emitters showed the opposite pattern, with elevated heart rate and slower respiration associated with decreased VOC emissions. This apparent reversal in coupling direction showed statistical significance in this exploratory sample for both heart rate (t=4.65, p=0.002, d=3.0) and respiratory rate (t=-4.93, p=0.0006, d=-2.8). These findings indicate that the direction of physiological-VOC coupling is not random but is systematically moderated by an individual's overall VOC emission level during stress. However, given the small subgroup sizes (n=6 per group), the extremely large effect sizes should be interpreted with caution and require replication in larger samples.

VI. STUDY 2B: MULTIMODAL CLASSIFICATION PERFORMANCE

*A. Feature Engineering*

Time-series data from all sensor modalities were segmented into 30-second non-overlapping windows to capture short-term physiological dynamics while maintaining sufficient temporal resolution for stress detection. The 30-second window size was chosen to balance between capturing transient stress responses, providing stable statistical estimates, and maintaining an adequate sample size for machine learning. For each window, we extracted the following features:

- Heart Rate: Mean, standard deviation, minimum, maximum, and range of heart rate values within each window (5 features)
- Galvanic Skin Response (GSR): Mean, standard deviation, minimum, maximum, and range of raw GSR conductance (5 features).
- Volatile Organic Compounds (VOC): Mean, standard deviation, minimum, maximum, range, temporal slope, and baseline-normalized mean of TVOC concentration in ppb (7 features from ENS160 TVOC sensor) and from the gas sensor BME688, we only used the 320 kΩ profile values. Mean, standard deviation, minimum, maximum, and range of gas resistance at 320°C in kΩ (5 features)

To account for inter-individual sensor calibration differences and focus on stress-induced relative changes, TVOC and Gas320 values were normalized as:

$$VOC_{norm} = \frac{VOC_t - VOC_{baseline}}{VOC_{Baseline}}$$

where $VOC_{baseline}$ represents the mean sensor reading during the 180-second baseline phase (Phase 2). This normalization yields dimensionless values representing proportional change from resting state, with positive values indicating VOC elevation. In total, 22 features were extracted per 30-second window. Feature extraction was performed only for Phases 2-7, excluding the initial questionnaire phase.

## B. Dataset Construction

Windows were labeled as "Stress" if they occurred during Phases 3-4 (stress-induction tasks) and "Non-Stress" if they occurred during Phases 2 and 5-7 (baseline and recovery). This yielded a dataset of 834 windows from 24 participants with a class distribution of 449 Stress (53.8%) and 385 Non-Stress (46.2%) samples. One participant was excluded due to insufficient data quality, resulting in 24 valid participants for analysis. Missing values in the feature matrix were imputed using column-wise mean imputation.

## C. Classification Models

We evaluated three supervised learning algorithms:

- Random Forest (RF): 100 trees, maximum depth of 10, using scikit-learn defaults for other hyperparameters

- Support Vector Machine with RBF kernel (SVM-RBF): C = 1.0, γ = 'scale'

- Support Vector Machine with Linear kernel (SVM-Linear): C = 1.0

## D. Cross-Validation Strategy

To evaluate model performance, we employed two complementary cross-validation approaches. First, stratified 5-fold cross-validation was conducted, with the dataset split into five folds stratified by class label to maintain equal proportions of stress/non-stress samples in each fold. Models were trained and evaluated 5 times, with each fold serving once as the test set, and performance metrics were averaged across folds with standard deviations reported. Second, to assess generalization to unseen individuals, we performed leave-one-subject-out cross-validation (LOSO), in which each of the 24 participants served once as the test set. At the same time, the model was trained on the remaining 23. This approach evaluates the model's ability to classify stress in individuals not represented in the training data, thereby providing a more conservative estimate of real-world generalization performance.

## E. Model Evaluation and Feature Analysis

To quantify the contribution of each sensor modality, we compared unimodal models trained separately on HR, GSR, TVOC (ENS160), and Gas320 (BME688) features with an early-fusion model trained on concatenated features from all modalities. Performance improvement was calculated as the difference in accuracy between early fusion and the best-performing unimodal model. To identify which features most strongly contributed to stress classification, we applied SHAP (SHapley Additive exPlanations) analysis to the Random Forest model. SHAP values quantify each feature's contribution to individual predictions based on Shapley values from cooperative game theory. We computed mean absolute SHAP values across all samples as a measure of global feature importance and aggregated importance by modality (HR, GSR, VOC) to assess relative contributions. Models were evaluated using accuracy (overall classification correctness), precision (proportion of true stress predictions among all stress predictions), recall (proportion of actual stress cases correctly identified), F1-score (harmonic mean of precision and recall), and AUC-ROC (area under the receiver operating characteristic curve).

## F. Results: RQ5 - Binary Stress Classification Performance

Table I presents the performance of three classification algorithms evaluated using stratified 5-fold cross-validation. Random Forest achieved the highest performance across all metrics, with an accuracy of 77.3% (±2.5%), F1-score of .789 (±0.025), and AUC of .847. Both SVM variants showed lower performance, with the RBF kernel (71.6% accuracy) outperforming the linear kernel (68.3% accuracy).

TABLE 2: CLASSIFICATION PERFORMANCE (5-FOLD CROSS-VALIDATION)

| Model | Accuracy | SD | Precision | Recall | F1-Score | SD | AUC |
|---|---|---|---|---|---|---|---|
| Random Forest | .773 | .025 | .791 | .786 | .789 | .025 | .847 |
| SVM (RBF) | .716 | .036 | .739 | .731 | .735 | .040 | .795 |
| SVM (Linear) | .683 | .050 | .714 | .688 | .701 | .056 | .765 |

*Note: SD = Standard Deviation across five folds. N = 834 windows from 24 participants.*

## G. Results: RQ6 - Generalization to Unseen Individuals

Leave-one-subject-out cross-validation revealed substantial inter-individual variability in classification performance. As shown in Table II, LOSO accuracy dropped to 65.3% (F1-score: 0.671), representing a 12 percentage point decrease compared to 5-fold CV. Individual participant accuracy ranged from 38.9% to 93.5%, indicating that some individuals exhibited highly predictable stress patterns while others showed greater ambiguity.

TABLE 3: LEAVE-ONE-SUBJECT-OUT CROSS-VALIDATION PERFORMANCE

| Metric | Value |
|---|---|
| Overall Accuracy | .653 |
| Overall F1-Score | .671 |
| Min Accuracy (Participant) | .389 |
| Max Accuracy (Participant) | .935 |
| Participants ($n$) | 24 |

*Note: Each participant served once as the test set with the remaining 23 as the training set.*

## H. Multimodal Fusion Analysis

Table III compares the performance of unimodal and multimodal classification approaches. Heart rate features alone achieved the highest unimodal performance (69.1% accuracy, AUC: .763), followed by TVOC features (66.2% accuracy, AUC: .669). GSR showed moderate discriminative power

(57.6% accuracy), while BME688 Gas320 performed near chance level (54.4% accuracy), likely due to high missing data rates. Early fusion of all modalities yielded 77.3% accuracy, representing a 12.0% relative improvement over the best unimodal approach (HR only).

TABLE 4: MULTIMODAL FUSION ANALYSIS

| Modality | N Features | Accuracy | F1-Score | AUC |
|---|---|---|---|---|
| HR+HRV | 5 | .691 | .714 | .763 |
| TVOC (ENS160) | 7 | .662 | .705 | .669 |
| GSR | 5 | .576 | .655 | .621 |
| Gas320 (BME688) | 5 | .544 | .691 | .504 |
| **Early Fusion (All)** | **22** | **.773** | **.789** | **.847** |

*Note: Improvement = +0.082 accuracy (+12.0%) vs. best unimodal (HR+HRV).*

### I. Feature Importance Analysis

An exploratory SHAP analysis suggested the relative contributions of each modality in this sample to stress classification decisions (see Table 4). Heart rate features dominated with 55.3% of total importance, with HR mean (0.058) and HR max (0.052) as the most influential individual features. VOC features contributed 24.9%, with normalized TVOC (0.028) ranking as the 6th most important feature overall. GSR features accounted for 19.8% of total importance.

TABLE 5: FEATURE IMPORTANCE BY MODALITY (SHAP ANALYSIS)

| Modality | Total Importance | Percentage | Top Features (Importance) | |
|---|---|---|---|---|
| HR | .233 | 55.3% | HR_mean<br>HR_max<br>HR_range | (.058)<br>(.052)<br>(.044) |
| VOC | .105 | 24.9% | TVOC_Norm<br>TVOC_min<br>TVOC_mean | (.028)<br>(.012)<br>(.011) |
| GSR | .084 | 19.8% | GSR_range<br>GSR_std<br>GSR_min | (.024)<br>(.019)<br>(.014) |

*Note: Importance values represent mean absolute SHAP values aggregated across 834 samples.*

Notably, normalized TVOC (TVOC_Norm_mean: .028) showed 2.5× higher importance than raw TVOC features (TVOC_ppb_mean: .011), indicating that baseline-referenced relative changes are more discriminative than absolute VOC concentrations for stress detection.

## VII. STUDY 2: DISCUSSION

### A. Integration of Physiological-VOC Coupling and Classification Performance

This exploratory empirical investigation provides foundational evidence that low-cost sensor setups can detect stress-related patterns through combined physiological and VOC measurements. Study 2a provided preliminary evidence that TVOC emissions may show structured, phase-dependent elevation during stress in high cardiovascular reactors. Study 2b demonstrated that these stress-related signals can be computationally leveraged for automated classification, achieving 77.3% accuracy through multimodal fusion.

The convergence of findings across both studies strengthens preliminary confidence for further investigation in VOC-based stress detection: Study 2a's physiological evidence (High reactors: $d$ = 1.38 TVOC increase) directly supports Study 2b's classification results (VOC features: 24.9% SHAP importance). The substantial interindividual heterogeneity identified in coupling analyses (CV>80%, coupling direction split 8:8) explains the LOSO performance degradation observed in classification (77.3% → 65.3%), as individual-specific response patterns require personalized calibration for optimal detection accuracy.

### B. Physiological-VOC Coupling Mechanisms

Study 2a revealed that cardiovascular reactivity moderates stress-induced TVOC changes, with High HR reactors showing substantial increases (d=1.38) while Low reactors exhibited no significant phase effects. This moderation likely reflects individual differences in sympathetic nervous system responsiveness rather than measurement artifacts, as the effect persisted after controlling for verbal load and respiratory patterns. However, the dissociation between trait-level group effects and absent moment-to-moment HR-TVOC correlations (r=0.086, p=0.697) within High reactors suggests complex, non-linear relationships between cardiovascular activity and VOC emission.

The observation that baseline emission levels appeared to moderate coupling direction (d=3.0 for HR; d=-2.8 for RR) in this exploratory sample represents a tentative finding requiring replication. These extremely large effect sizes based on small subgroups (n=6 each) likely reflect sampling variability and should not be generalized without validation in adequately powered studies (n≥50). High emitters may have metabolic profiles or pulmonary gas-exchange characteristics that amplify VOC release during sympathetic activation, whereas low emitters may exhibit VOC sequestration or altered clearance kinetics during stress. Alternatively, sensor proximity effects could differ based on baseline emission concentrations, with high emitters saturating nearby sensors more rapidly. The optimal temporal lags of 30-80 seconds align with expected pulmonary transit times and peripheral-to-central circulation delays, supporting physiological plausibility.

*C. Multimodal Classification Architecture*

Study 2b demonstrated that multimodal fusion combining HR, GSR, and VOC sensors achieves 12% accuracy improvement over the best unimodal approach (HR alone: 69.1%; early fusion: 77.3%). This complementarity likely reflects different temporal dynamics and physiological mechanisms: heart rate responds rapidly to sympathetic activation (seconds), GSR reflects sudomotor nerve activity with moderate latency (1-3 seconds), and VOC changes may capture metabolic shifts occurring over longer timescales (30-80 seconds per Study 2a coupling analyses).

SHAP analysis revealed that while physiological signals (HR, GSR) contributed 75% of classification performance, VOC sensors provided substantial complementary information (25%) despite being non-invasive breath-based measurements. The 2.5× higher importance of normalized versus raw TVOC features confirms Study 2a's emphasis on individual calibration: baseline-referenced relative changes capture stress reactivity more effectively than absolute concentrations.

The poor performance of BME688 Gas320 features (54.4% accuracy, near chance) contrasts with the moderate success of ENS160 TVOC features (66.2% accuracy). This disparity may reflect sensor-specific differences in target compound selectivity, with ENS160's broader TVOC sensitivity better suited for capturing complex stress-related volatile mixtures compared to BME688's temperature-cycled resistance measurements. Alternatively, the 30-second windowing may not optimally capture BME688's multi-temperature cycling dynamics (250°C, 320°C, 400°C with 150ms heating and 250ms stabilization delays).

*C. Individual Differences and Generalization Challenges*

The 12 percentage point LOSO performance drop (77.3% → 65.3%) and wide per-participant accuracy range (38.9%-93.5%) reflects substantial inter-individual response heterogeneity identified in Study 2a. The even split in coupling direction (8 positive, 8 negative correlations) among responders explains why population-level models show reduced generalization: individuals with opposing physiological-VOC relationships contribute conflicting signal patterns during training.

These exploratory findings suggest a fundamental tension: the need for personalized models to achieve high accuracy versus the desire for generalizable population-level algorithms. The moderate LOSO performance (65.3%) suggests that while a universal model captures broad stress patterns, personalized calibration using even small amounts of individual-specific data could improve real-world performance. The baseline emission moderator effect (d=3.0) identified in Study 2a provides a potential stratification criterion: clustering participants by baseline TVOC levels during initial calibration could enable group-specific models with improved generalization.

*D. Protocol Standardization Requirements*

The convergence of Study 2a's coupling heterogeneity and Study 2b's generalization challenges underscores critical needs for methodological standardization. Sensor-to-mouth distance (5-6 cm in this study) affects TVOC concentration measurements, as exhaled breath disperses rapidly in ambient air. Verbal load effects require systematic control: recovery phases with speech (counting) versus silence showed non-significant TVOC differences ($p = .084$), but larger samples may detect subtle confounds. BME688 temperature cycling profiles (250°C, 320°C, 400°C) require optimization, as different temperatures target distinct VOC classes (alcohols, aldehydes, alkanes). Environmental controls, including room ventilation protocols, ambient VOC baselines, and humidity/temperature stability, all require harmonization across studies.

The 30-second window size used in Study 2b represents a pragmatic compromise but may not optimally capture all physiological dynamics. Heart rate variability analysis typically benefits from longer windows (≥5 minutes), whereas TVOC temporal slopes were important for classification (6th-ranked feature), suggesting that finer temporal resolution could improve performance. Future work should systematically evaluate window sizes from 15 seconds to 5 minutes to identify optimal trade-offs between temporal resolution, feature stability, and sample size.

*E. Limitations and Interpretative Caution*

The small sample ($n = 25$) and even smaller responder subgroups ($n = 6$-$8$) represent the primary limitation of this exploratory investigation. The extremely large effect sizes observed for baseline emission moderators ($d = 3.0$ for HR coupling, $d = -2.8$ for RR coupling) should be interpreted with extreme caution, as they likely reflect sampling variability rather than true population effects. Post-hoc sensitivity analysis ($\alpha = .05$, power $= .80$) indicates our design could reliably detect only very large effects ($d > 1.2$) in subgroup comparisons. We recommend $n \geq 50$ for future confirmatory studies, with $n \geq 15$ participants per stratified subgroup to enable adequately powered moderator analyses.

The exploratory nature of this investigation entails several important limitations affecting both Study 2a and 2b. The small sample size ($n = 25$, reduced to n=24 for classification) limits generalizability, with Study 2a responder subgroup analyses restricted to n=6-8 participants. The extremely large effect sizes observed for baseline emission moderators (d=3.0 for HR coupling, d=-2.8 for RR coupling) should be interpreted with extreme caution, given these small subgroups, and require replication in larger samples before drawing firm conclusions.

Equipment failures resulted in incomplete data collection (GSR $n = 19$; EEG $n = 2$ only), preventing a comprehensive multimodal physiological assessment. Measurements across three different laboratory spaces introduced environmental heterogeneity that may have contributed to inter-individual variability observed in both coupling patterns and classification performance. The single-session design prevented test-retest reliability assessment, leaving it unknown whether observed coupling patterns and classification accuracies represent stable traits or state-dependent phenomena.

Laboratory-only conditions limit ecological validity, as real-world stress may elicit different VOC response profiles. The ENS160 sensor provides only total VOC concentration without compound-specific identification, precluding comparison with GC-MS-identified biomarkers from Study 1's systematic review. Despite pre-session instructions, confound control compliance (e.g., food intake, fragrance use) was not biochemically verified. Recovery-phase data (6 minutes) remain underanalyzed, and only cognitive/evaluative stress was tested, limiting generalizability to other stressors, such as social threat or physical exertion. The 77.3% classification accuracy achieved in Study 2b, while encouraging for proof-of-concept, remains insufficient for clinical diagnostic applications. The 65.3% LOSO accuracy represents acceptable but reduced generalization, requiring personalized calibration for deployment. Given these limitations, findings should be considered preliminary evidence requiring replication and extension before informing applied affective computing systems.

*G. Implications for Affective Computing*

The preliminary feasibility demonstrated for combining non-invasive VOC sensors with traditional physiological measurements opens new avenues for unobtrusive stress monitoring in naturalistic settings. Unlike chest-worn heart rate monitors or finger-placed GSR electrodes, breath sampling could potentially be integrated into wearable devices (e.g., face masks, collar-mounted sensors) with minimal user burden. However, Study 2a's findings regarding individual calibration requirements and Study 2b's LOSO performance degradation indicate that practical deployment would require brief baseline collection periods (e.g., the 180-second protocol used here) before reliable stress detection.

The preliminary convergence observed between physiological patterns (Study 2a) and classification performance (Study 2b) in this exploratory sample tentatively suggests that VOC-based sensing provides complementary information to established modalities rather than serving as a standalone replacement. The 24.9% SHAP importance of VOC features indicates meaningful contribution beyond traditional sensors, justifying the added complexity of multi-sensor systems. However, the baseline emission moderator effect (d=3.0) and coupling direction heterogeneity (8:8 split) indicate that population-level thresholds will fail; individualized models or stratified approaches are essential.

For real-world applications such as workplace wellness monitoring, biofeedback training, or adaptive human-computer interfaces, the combination of Study 2a's mechanistic insights and Study 2b's classification architecture provides a roadmap: (1) initial calibration sessions to establish individual baseline TVOC levels and coupling patterns, (2) stratification by baseline emission levels to select appropriate population submodels, (3) multimodal fusion emphasizing baseline-normalized features, and (4) periodic recalibration to account for temporal drift in sensor response or physiological characteristics.

ACKNOWLEDGMENT

The authors gratefully acknowledge Volker Wunsch from Mannheim University of Applied Sciences for his technical consultation and support.

ETHICAL APPROVAL

All procedures performed in studies involving human participants were in accordance with the ethical standards of the institutional and/or national research committee and with the 1964 Helsinki declaration and its later amendments or comparable ethical standards.